\documentclass[fleqn,usenatbib]{mnras}

\usepackage{newtxtext,newtxmath}
\usepackage[T1]{fontenc}

\DeclareRobustCommand{\VAN}[3]{#2}
\let\VANthebibliography\thebibliography
\def\thebibliography{\DeclareRobustCommand{\VAN}[3]{##3}\VANthebibliography}

\usepackage{graphicx}	% Including figure files
\usepackage{amsmath}	% Advanced maths commands

\usepackage{orcidlink}

\usepackage{caption}
\usepackage{subcaption}

\title[Composition and winds of TOI-1518 b]{PEPSI Investigation, Retrieval, and Atlas of Numerous Giant Atmospheres (PIRANGA). III. Composition and winds in the atmosphere of TOI-1518~b}

\author[Basinger et al.]{
Connor Basinger,$^{1}$\thanks{E-mail: basinger.101@osu.edu}
Marshall C. Johnson \orcidlink{0000-0002-5099-8185},$^{1}$
Ji Wang \orcidlink{0000-0002-4361-8885},$^{1}$
Alison Duck \orcidlink{0000-0002-4531-6899},$^{1}$
Anusha Pai Asnodkar \orcidlink{0000-0002-8823-8237},$^{1}$\newauthor
Sydney Petz,$^{1}$
Calder Lenhart,$^{1}$
Ilya Ilyin$^{2}$
and Klaus Strassmeier$^{2,3}$
\\
$^{1}$Department of Astronomy, The Ohio State University, 140 West 18th Avenue, Columbus, OH 43210, USA\\
$^{2}$Leibniz-Institute for Astrophysics Potsdam (AIP), An der Sternwarte 16, D-14482 Potsdam, Germany\\
$^{3}$Institute of Physics \& Astronomy, University of Potsdam, Karl-Liebknecht-Str. 24/25, D-14476 Potsdam, Germany\\
}

\date{Accepted XXX. Received YYY; in original form ZZZ}

\pubyear{2025}

\begin{document}
\label{firstpage}
\pagerange{\pageref{firstpage}--\pageref{lastpage}}
\maketitle

\begin{abstract}
Ultra-hot Jupiters (UHJs) orbit close to their host stars and experience extreme conditions, making them important laboratories to explore atmospheric composition and dynamics. 
Transmission spectroscopy is a useful tool to reveal chemical species and their vertical and longitudinal distribution in the atmosphere.
We use transmission spectra from the PEPSI spectrograph on the Large Binocular Telescope to search for species and measure their time-resolved wind velocities in the atmosphere of TOI-1518~b. We detect Fe I at 7.8$\sigma$ and Fe II at 8.9$\sigma$, and tentatively detect Cr I at 4.4$\sigma$ and Ni I at 4.0$\sigma$. The time-resolved wind velocities of Fe I show a velocity pattern that is consistent with the velocity pattern of Fe II.
TOI-1518~b joins a small sample of UHJs for which time-resolved wind velocities have been measured.
\end{abstract}


%%%%%%%%%%%%%%%%%%%%%%%%%%%%%%%%%%%%%%%%%%%%%%%%%%

%%%%%%%%%%%%%%%%% BODY OF PAPER %%%%%%%%%%%%%%%%%%

\section{Introduction}

Ultra-hot Jupiters (UHJs) are important objects for atmospheric study due to their close orbits to their host stars, large radii, and high temperatures. Transmission spectroscopy is a powerful tool for this research, enabling the detection of atmospheric species by searching for their absorption lines in the terminator, i.e. the transition region between the day and night sides of the exoplanet. UHJs are particularly favorable targets for transmission spectroscopy because their high equilibrium temperatures directly lead to large atmospheric scale heights, which provides high signal-to-noise ratio (SNR) data. This high-quality data is crucial for exploring the extreme atmospheric properties of UHJs including large day-to-night temperature contrasts which drive strong day-to-night winds (e.g. \citealt{Snellen2010,Ehrenreich2020,PaiAsnodkar2022}), super-rotating equatorial jets (\citealt{Showman2009,Showman2011}), or high dayside temperatures which lead to extreme chemistry (\citealt{Cooper2006, Bell2018, Showman2020}). It is also possible to distinguish the spatial locations of different atmospheric species, which can further aid in determining the properties and circulation patterns of the atmosphere. Detecting and cataloging additional atmospheric species is crucial to evaluating the chemistry of UHJ atmospheres. Furthermore, the frequent transits of UHJs offer flexibility in obtaining observations and enable the repeated measurements needed to verify these atmospheric properties.

Among $\sim$19 UHJs with equilibrium temperatures $T_{eq}>2000$ K, radius $R>R_{J}$, and magnitude $V < 10$, (e.g. \citealt{West2016,Gaudi2017,Evans2017}), 
TOI-1518~b is an interesting target for several reasons. First, the discovery paper found that the planet has a near-grazing transit with an impact parameter b$\sim$0.9~\citep{Cabot2021}. They also detected Fe I and found evidence for Fe II in the transmission spectra and suggested that it may be a future target for emission spectroscopy to search for thermal inversions, where the atmospheric temperature increases with altitude. A search for a thermal inversion was later completed in \cite{Petz2024PIRANGA}, who used emission spectroscopy to confirm the existence of an inversion. The agent responsible for thermal inversions remains elusive. While molecules like TiO and VO are often suspected to be responsible \citep{Hubeny2003,Fortney2008} for thermal inversions, they are very rarely detected \citep{Nugroho2020,Prinoth2022,Johnson2023}. They are likely photodissociated and therefore do not play an important role. There is increasing evidence that other metals, particularly Fe, which is universally detected in UHJ atmospheres, absorb in the ultraviolet (UV) and optical, offering a compelling alternative explanation to the inversion phenomena \citep{Lothringer2018,Gandhi2019,Petz2024PIRANGA,Chachan2025}.

Another reason making TOI-1518~b an interesting target is that~\citet{Watanabe2024} found evidence for nodal precession of the planetary orbit, making it the fourth system to have evidence of orbital nodal precession \citep{Szabo2012,Johnson2015,Stephan2022}. They found that the impact parameter is changing at a rate of $db/dt = -0.0116 \pm 0.0036 \text{ yr}^{-1}$, meaning that the transit is currently moving toward the center of the star and will take well over a century until it is no longer visible. This occurs as a result of the oblateness of the host star due to its rapid rotation and the nearly polar orbit of the exoplanet. Nodal precession changes the transit parameters needed to model the Doppler shadow.

The Doppler shadow is a dark shaded region which appears on a cross-correlation map. It appears because, during the transit, the exoplanet blocks flux from local regions of the host star which are blueshifted or redshifted. The Doppler shadow's shape traces out the path the exoplanet takes across the host star. The Doppler shadow helps to confirm transiting planets around fast-rotating stars \citep{Gaudi2007,CollierCameron2010}, measure spin-orbit (mis)alignment \citep{CollierCameron2010,Johnson2017}, and monitor long-term nodal precession \citep{Johnson2015,Watanabe2024}. The Doppler shadow can also contaminate atmospheric data by overlapping the atmospheric signal (e.g. \citealt{CasasayasBarris2022}) and overlaps the atmospheric signal of TOI-1518~b near ingress \citep{Cabot2021}, but can be modeled and removed to facilitate the detection of species in the atmosphere.

Moreover, the proximity of UHJs to their host stars leads to significant rotation during transit (e.g., $\sim 30 ^\circ$ in the case of WASP-76~b; \citealt{Ehrenreich2020}). This reveals different regions of the exoplanet's atmosphere which can be probed with a series of high-resolution transmission spectra, or time-resolved spectroscopy. 
\cite{Ehrenreich2020} observed a progressively increasing blueshift for Fe I in WASP-76~b until it reached a roughly constant velocity for the remainder of the transit. They attributed this to first observing the leading edge transit the host star, revealing the morning side of the terminator where the velocity of the blueshifted day-to-night winds is partially offset by the exoplanet's rotation and equatorial jet. Fe I was expected to condense out on the cooler nightside and may largely be absent from the morning terminator. As the transit progressed, the evening side of the terminator with blueshifted Fe I dominated the signal. The evening terminator is expected to be hotter than the morning terminator due to an offset in the hottest spot on the exoplanet's dayside, so Fe I condensation on the evening terminator should be less significant.
Similar behavior with an increasing blueshift over time was shown for WASP-121~b \citep{Borsa2021}.
Alternative explanations to explain this wind velocity pattern have been proposed. For example, in \cite{Wardenier2021} the large temperature difference between the morning and evening terminators could reproduce this velocity pattern even absent any iron condensation. 
In \cite{Savel2022}, the velocity pattern could not be reproduced in their general circulation models (GCMs) with iron condensation alone. Instead, the inclusion of high-altitude, optically thick clouds and a small orbital eccentricity best reproduced the velocity pattern.

In this paper, we utilize high-resolution spectroscopic data from the PEPSI spectrograph at the Large Binocular Telescope (LBT) to (1) search for evidence of atomic and molecular species, (2) conduct time-resolved wind velocity measurements in the transmission spectra of TOI-1518~b, and (3) measure the transit parameters of TOI-1518b for preliminary comparison with the expected values and previously detected nodal precession. In \S2, we describe our high-resolution spectroscopic observations. \S3 describes our data reduction, telluric removal, and systematics removal. In \S4, we describe our template spectra, cross-correlation functions, and Doppler shadow removal. Finally, \S5 presents and discusses our detected species and measured wind velocities.

\section{Observations}

Data were obtained on 2022 September 26 UT using the Potsdam Echelle Polarimetric and Spectroscopic Instrument (PEPSI; \citealt{Strassmeier2015}) on the LBT. The LBT has dual 8.4m apertures with the light simultaneously fed from both primary mirrors to the spectrograph. Two different wavelength ranges were observed simultaneously: 4800-5441{\AA} on the blue arm (CD3) and 6278-7419{\AA} on the red arm (CD5). The resolving power of both is R$\sim$130,000 with a fiber diameter of 200 microns. A series of 50 exposures were taken over the course of the night, before, during, and after the transit of TOI-1518~b. Each observation had an exposure time of 250s with a median continuum SNR of 146 (blue arm) and 181 (red arm). 

Several exposures near ingress were lost due to high extinction from clouds, causing the instrument to lose the guide star. We expect this to reduce our detection strength slightly. Due to the transit geometry, these are also among the observations in which the Doppler shadow and atmospheric signal overlap, possibly reducing the importance of removing the Doppler shadow (\S4). It also means there were fewer observations that can be incorporated into phase bins during ingress. This decreases the signal on the time-resolved wind velocities during ingress compared to egress and we dynamically increase the bin size to achieve a higher SNR (\S5). 

\section{Data Reduction and Processing}

Data reduction was performed via the Spectroscopic Data Systems (SDS) pipeline (see \citealt{Ilyin2000} and \citealt{Strassmeier2018}). 
The process includes bias correction by subtracting a smoothed spline function fit to the overscan regions, 
photon noise estimation for the variance, 
flat-field correction using master flats, scattered light removal using a smoothed spline function fit to the gaps between diffraction orders, 
spectral orders defined by fitting an elongated Gaussian to each slice of each order, 
maximizing SNR by normalizing flux and rejecting cosmic rays, 
calibrating wavelengths by matching emission lines to a wavelength table and correcting for barycentric motion, 
correcting the spectra for the blaze function, vignetting, and fringing, 
and finally fitting the continuum to the extracted spectral orders.

\begin{figure}
  \centering
	\includegraphics[width=\columnwidth]{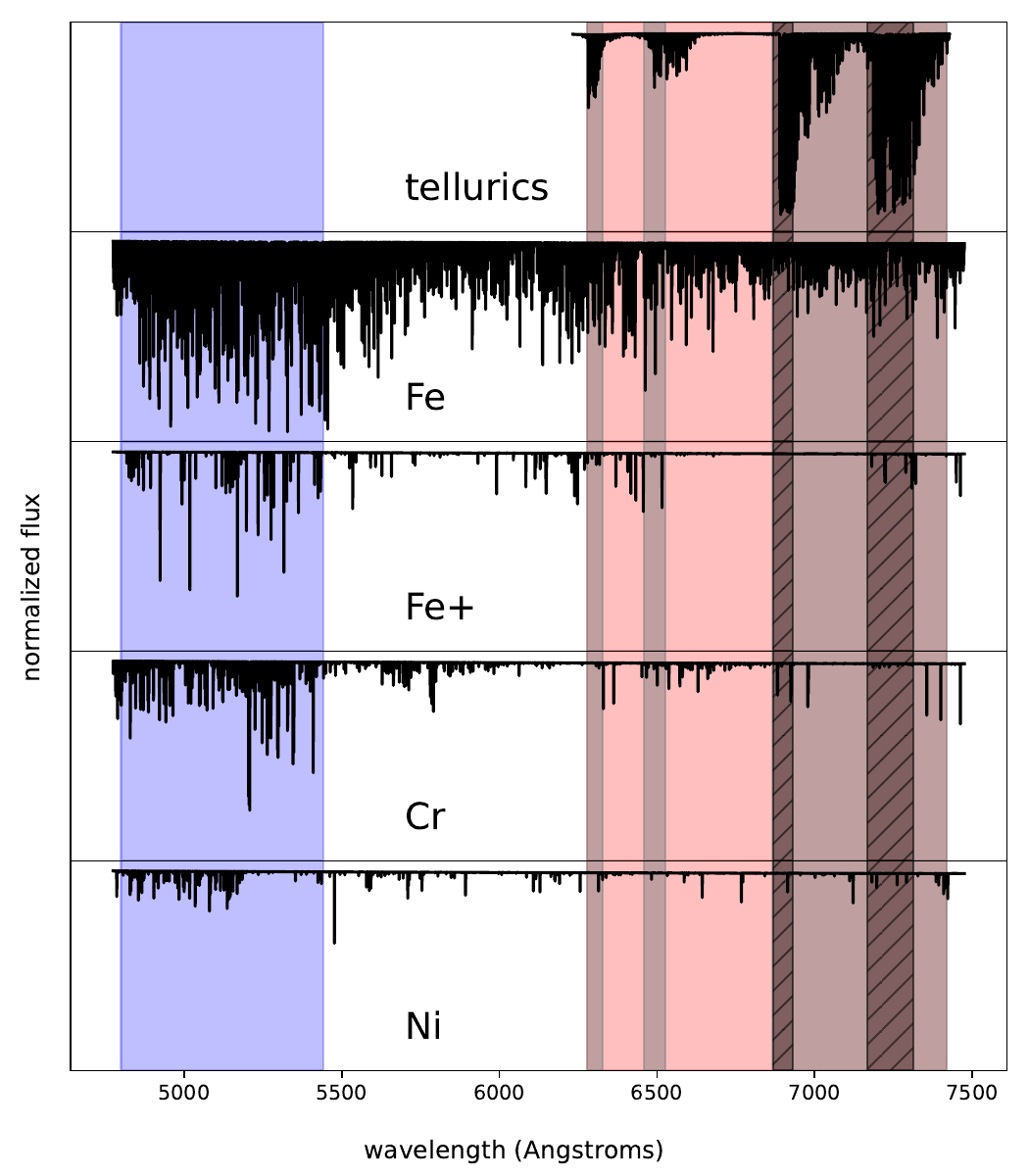}
    \caption{(Top) Model telluric spectrum produced by MOLECFIT, scaled to fit on the same y-axis as the template spectra. (Other panels) Template transmission spectra for TOI-1518~b assuming a single species and solar H and He in its atmosphere. The VMRs (Table~\ref{tbl:species}) were determined via a grid search that maximized SNR. The y-axis for each species is on the same scale. The blue shaded region corresponds to the wavelengths observed by the PEPSI blue arm and the red shaded regions correspond to the wavelengths observed by the PEPSI red arm. The grey shaded region identifies wavelengths that have significant telluric contamination, while the hatched regions are masked out entirely in our analysis due to the presence of strong telluric contamination.}
    \label{fig:TempFeI}
\end{figure}

We fit and removed telluric lines from the red arm using Molecfit (\citealt{Kausch2015}; \citealt{Smette2015}). The tellurics affecting the data from the red arm are shown in the top panel of Fig.~\ref{fig:TempFeI}. The blue arm is largely telluric-free, with only weak ozone absorption bands present. These are removed via the implementation of SYSREM below.

During a transit, our spectra are the combination of the host star's spectrum and the transmission spectrum from the light passing through the atmosphere of the exoplanet. 
The exoplanet's spectrum shifts significantly in velocity space over the course of the transit, while the host star's and telluric spectra do not. We calculated the median spectrum based on all observations. This removes the exoplanetary spectra due to their velocity shifts and leaves the signal of the host star. We subtracted the median from the raw data to obtain the transmission spectrum of TOI-1518~b. The choice of subtracting or dividing by the median spectrum does not significantly affect the residual data or our analysis.

Following the data treatment procedures in \cite{Johnson2023}, we implement SYSREM (\citealt{Tamuz2005}) to remove systematics from the data before computing the cross-correlation function (CCF). Our implementation is a modified version of PySysRem \footnote{\href{https://github.com/stephtdouglas/PySysRem}{https://github.com/stephtdouglas/PySysRem}} with formal error propagation through the algorithm. The guess for the initial iteration of SYSREM is based on the airmass. For TiO, VO, FeH, and CaH, we remove the adopted number of systematics listed in Table 5 of \cite{Johnson2023}. For other species, we remove 3 iterations in the blue arm and 1 in the red arm for spectral ranges with no tellurics. In ranges of the red arm that have tellurics we remove 10 systematics.

\section{Analysis}

Cross-correlating a template spectrum with the residual data may be used to search for the presence of species in the transmission spectrum \citep{Snellen2010,Brogi2012,Birkby2018}. To calculate template spectra for the different species, we used petitRADTRANS \citep{Molliere2019}, a Python package that models transmission spectra given input conditions of the atmosphere (e.g. a pressure-temperature profile) and the mass fractions of the elements that are present in the atmosphere. Each template spectrum typically contains a single species to enable us to search for that species in the exoplanet's atmosphere. Sources for the high-resolution opacities can be found in a table in the petitRADTRANS documentation for available opacity species\footnote{\href{https://petitradtrans.readthedocs.io/en/latest/content/available_opacities.html}{https://petitradtrans.readthedocs.io/en/latest/content/available\_opacities.html}}. Other opacities not found in this table were taken from the DACE opacity database\footnote{\href{https://dace.unige.ch/opacityDatabase/}{https://dace.unige.ch/opacityDatabase/}} and converted to a format readable by petitRADTRANS (\citealt{Petz2024PETS} and Molli{\`e}re, private communication). We adopted a Guillot pressure-temperature profile (\citealt{Guillot2010}) with $\gamma = 45$ and $\kappa_{IR} = 0.4$ \citep{Petz2024PIRANGA}. The abundances of H and He on the exoplanet are assumed to be solar with a log relative abundance A(He) = 10.925 on a scale where hydrogen is 12. We adopted a constant mean molecular weight (MMW) of 2.33. Choosing to instead calculate the MMW throughout the atmosphere and comparing the resulting spectrum to the template generated using a constant MMW results in fractional changes at the level of less than $1 \times 10^{-4}$. 

The volume mixing ratio (VMR) is related to the mass fraction by $X_i = \mu_i/\mu * n_i$, where $X_i$ is mass fraction, $\mu_i$ is the species mass, $\mu$ is the MMW, and $n_i$ is the VMR. The VMR is selected for the different species by performing a grid search of VMRs to maximize the SNR of each of our species. This is different than using chemical equilibrium models such as FastChem \citep{Stock2018}, which was used, e.g., for KELT-20b in \cite{Petz2024PETS}. Chemical equilibrium models may underproduce abundances of some elements. For example, the models neglect the contribution of photoionization from the host star's radiation and underproduce Fe II. Our method avoids unrealistic low-abundances for certain species of atoms and molecules (e.g., Fe II) so that the absorption lines generated in the model are always significant and robust for cross-correlation. We caution that this method should not be used as a determination of the absolute abundances of the species we detect in this work. This is sufficient for simply detecting the presence of atoms and molecules, but would not be satisfactory for a more in-depth study with the goal of determining their abundances. 

The transmission spectrum generated by petitRADTRANS is given in transit radii. For comparison to the observed spectra, this was converted to normalized flux by 
$F = 1 - \left[ \left( \frac{R_p}{R_\star} \right)^2- \left( \frac{R_0}{R_\star} \right)^2 \right]$ \citep{Johnson2023}, 
where $R_p$ is the wavelength-dependent transit radius, $R_0 = 1.875 R_J$ is the planetary baseline radius from the photometric analysis in \citet{Cabot2021}, and $R_\star$ is the stellar radius. 
Fig.~\ref{fig:TempFeI} shows the template transmission spectra for each individual atmosphere with Fe I, Fe II, Cr I, and Ni I. The y-axis for each species is on the same scale. 

\begin{figure}
  \centering
	\includegraphics[width=\columnwidth]{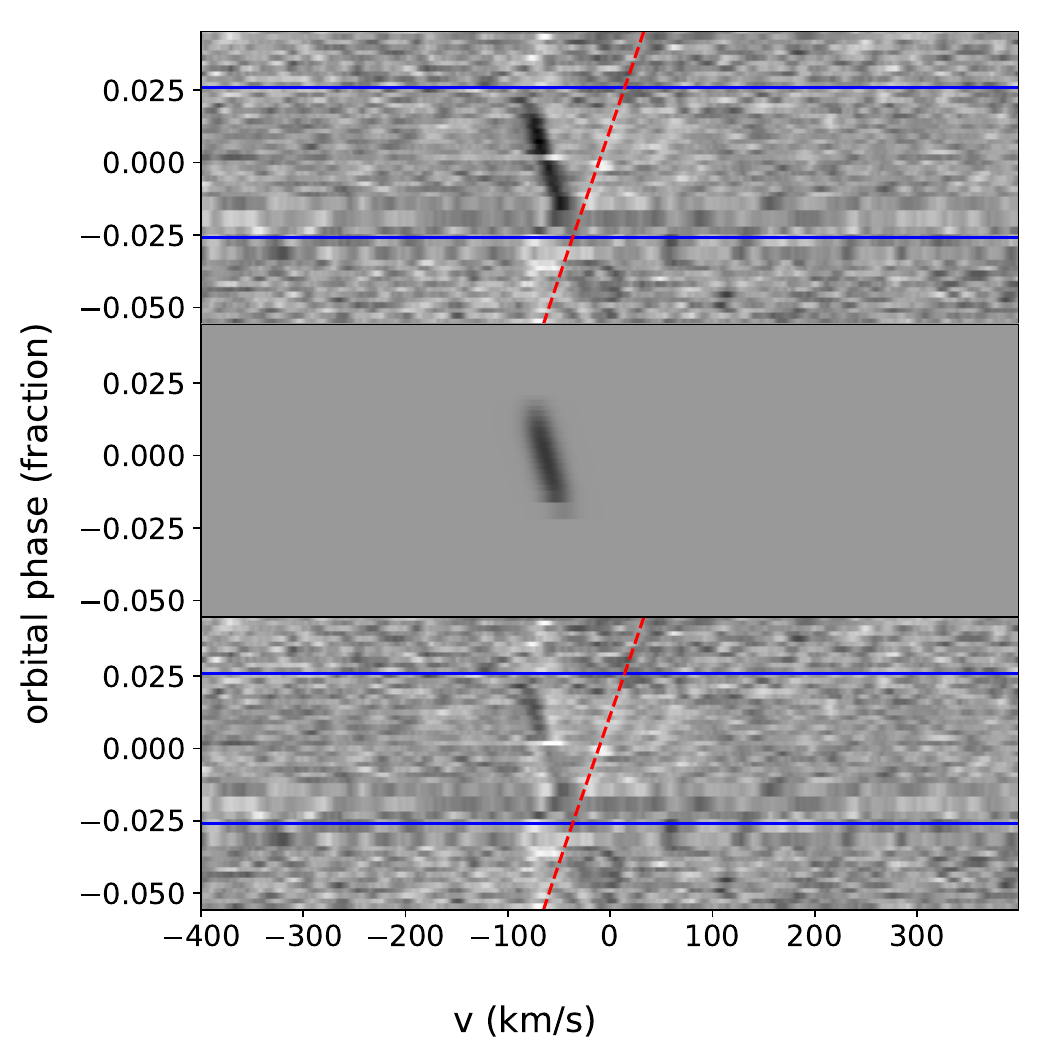}
    \caption{(Top) Raw CCF for Fe I. Dark (black) regions are anti-correlated and light (white) regions are correlated. 
    The blue lines show the ingress and egress of the transit, calculated using the transit duration determined by Duck et al. (in prep). The red dotted line shows the expected velocity of the atmospheric signal $V_{sig} = K_P \sin(2 \pi \phi) + V_{sys}$, where $K_P = 157$ km s$^{-1}$ is the radial velocity semi-amplitude (\citealt{Cabot2021}), $\phi$ is the orbital phase, and $V_{sys} = -11.17 \pm 0.035$ km s$^{-1}$ is the systemic velocity offset in the LBT PEPSI frame \citep{Petz2024PIRANGA}.
    The dark shaded region is the Doppler shadow. It makes a V shape with the light shaded region, which is the observed atmospheric signal.
    (Middle) Model of the Doppler shadow from HORUS. This adopts transit parameters shown in Table~\ref{tbl:mcmcparam} which were obtained via MCMC with MISTTBORN.
    (Bottom) Residual after subtracting the Doppler shadow.}
    \label{fig:CCFstack}
\end{figure}

The CCFs are calculated by shifting the template spectrum through a range of Doppler shifts and calculating the dot product between the shifted template spectrum and the transmission spectrum. The velocity range used here was -400 to 400 km s$^{-1}$ with a spacing of 1 km s$^{-1}$. The raw 2D CCFs (i.e., CCF vs. orbital phase) are shown in the top panel of Fig.~\ref{fig:CCFstack}. Each CCF has been divided by the variance to obtain the SNR, where the variance is equal to the quadrature sum of the standard deviation of the residuals and the systematic error calculated by SYSREM. Lighter colors are correlated,
and the white track marked by the red dotted line is the absorption signal.

The dark shaded region is the Doppler shadow, which results from the Rossiter-McLaughlin effect \citep{Rossiter1924, McLaughlin1924}. The Rossiter-Mclaughlin effect occurs because the host star is rotating, so local points on the host star will be moving towards or away from the observer and undergo a Doppler shift. The transiting exoplanet blocks only a local region of the host star, so its flux from that region is not observed in the raw data. The Doppler shadow then traces the radial velocities of the patches of the host star that have been blocked by the exoplanet. The Doppler shadow and the planet atmospheric absorption signal should both be considered when modeling the observation data. We modeled and removed the Doppler shadow from the CCF before analyzing the atmospheric signal. 

To compute our model of the Doppler shadow, we first determine the transit parameters using MISTTBORN\footnote{\href{https://github.com/captain-exoplanet/misttborn}{https://github.com/captain-exoplanet/misttborn}} and HORUS. MISTTBORN allows fitting of planetary parameters using MCMC methods (using the emcee package, \citealt{ForemanMackey2013}) and HORUS. HORUS is used for Doppler tomographic modeling. See \S3.3 of \cite{Johnson2014} or \S2.2 of \cite{Johnson2017} for additional details. 
An MCMC analysis is used to reproduce the Doppler shadow seen in Fig. 2, solving for parameters including $\lambda$, b, and v sin i, the projected obliquity, impact parameter, and projected rotation speed, respectively. The model and priors from the literature are combined into a single $\chi^2$-like value for each step in the chain. 
We placed priors on the period, transit epoch, ratio of semi-major axis to stellar radius, and ratio of planetary to stellar radius because they are better constrained via photometric datasets with many transits than the single transit in our observations. The adopted priors are shown in Table~\ref{tbl:mcmcparam}. MISTTBORN does not currently support asymmetric priors. 
The model of the Doppler shadow is shown in the middle panel of Fig.~\ref{fig:CCFstack}. The results of the MCMC simulation are summarized in Table~\ref{tbl:mcmcparam}. 

\begin{table*} 
\caption{Parameters of TOI-1518 from MISTTBORN. Priors are given by either a uniform ($\mathcal{U}$) or normal ($\mathcal{N}$) distribution. *We introduced a scale factor to MISTTBORN to account for the depth of the Doppler shadow compared to the input CCFs. The input CCFs are in SNR space. Dividing these by the peak SNR multiplied by some factor yields a Doppler shadow $\sim$1\% the depth of the normalized line profile. This scale factor also alters the reported $R_P/R_{\star}$, so the value we quote here is not directly comparable to $R_P/R_{\star}$ from the literature. ** $k = R_P/R_\star$ ***To avoid our model choosing an anomalously high v sin i, which would not be supported by independent constraints of the star's rotation, we reduced the uncertainty on the prior by a factor of 100. This factor was determined by reducing the uncertainty by an order of magnitude until the model consistently adopted v sin i roughly equal to the literature value.} 
\begin{tabular}{llccc} 
\hline 
\hline 
Description & Parameter & Value & Prior & Prior Source\\ 
\hline 
Measured Parameters \\ 
Transit epoch & $T_0$ (BJD) & $2459882.95358^{+0.00014}_{-0.00011}$ & $\mathcal{N}(2459882.953414,9.4\times10^{-5})$ & Duck et al. (in prep) \\ 
Orbital Period & $P$ (days) & $1.902615^{+1.2\times10^{-5}}_{-1.1\times10^{-5}}$ & $\mathcal{N}(1.902605, 6.5\times10^{-6})$ & Duck et al. (in prep) \\ 
Scaled* ratio of planetary radius to stellar radius & $R_P/R_{\star}$* & $0.07531^{+0.00074}_{-0.00055}$ & $\mathcal{U}(0,1)$ & This Work \\ 
Impact parameter & $b$ & $0.9019^{+0.0011}_{-0.0058}$ & $\mathcal{U}(-(1+k),1+k)$** & This Work\\ 
Ratio of planetary semi-major axis to stellar radius & $a/R_{\star}$ & $4.066^{+0.05}_{-0.052}$ & $\mathcal{N}(4.128, 0.028)$ & Duck et al. (in prep)\\ 
Projected rotation speed & $v\sin i_{\star}$ (km s$^{-1}$) & $74.502^{+0.035}_{-0.034}$ & $\mathcal{N}(74.4, 0.023)$*** & \cite{Cabot2021} \\ 
Projected obliquity & $\lambda$ ($^{\circ}$) & $246.44^{+0.65}_{-0.43}$ & $\mathcal{U}(-\infty,\infty)$ & This Work \\ 
Limb darkening coefficient & $q_{1,\mathrm{tom}}$ & $0.5129^{+0.0106}_{-0.0094}$ & $\mathcal{N}(0.4469, 0.01)$ & \cite{Claret2013}\\ 
Limb darkening coefficient & $q_{2,\mathrm{tom}}$ & $0.309^{+0.023}_{-0.011}$ & $\mathcal{N}(0.2193, 0.01)$ & \cite{Claret2013}\\ 
Width of intrinsic stellar line profile & $v_{\mathrm{int}}$ (km s$^{-1}$) & $13.07^{+0.24}_{-0.2}$ & $\mathcal{U}(0,\infty)$ & This Work\\ 
\hline 
Derived Parameters \\ 
& $i$ ($^{\circ}$) & $77.253^{+0.095}_{-0.3}$ &  & \\ 

\hline 
\end{tabular} 
\label{tbl:mcmcparam}
\end{table*}

Subtracting the Doppler shadow from the line profile residuals reveals more clearly the atmospheric signal, a bright streak close to the red dotted line. Since the Fe I template was used here in cross-correlation, this indicates the presence of Fe I in the exoplanet's atmosphere (bottom panel of Fig.~\ref{fig:CCFstack}). 

\begin{figure*}
     \centering
     \begin{subfigure}[b]{\columnwidth}
         \centering
         \includegraphics[width=\textwidth]{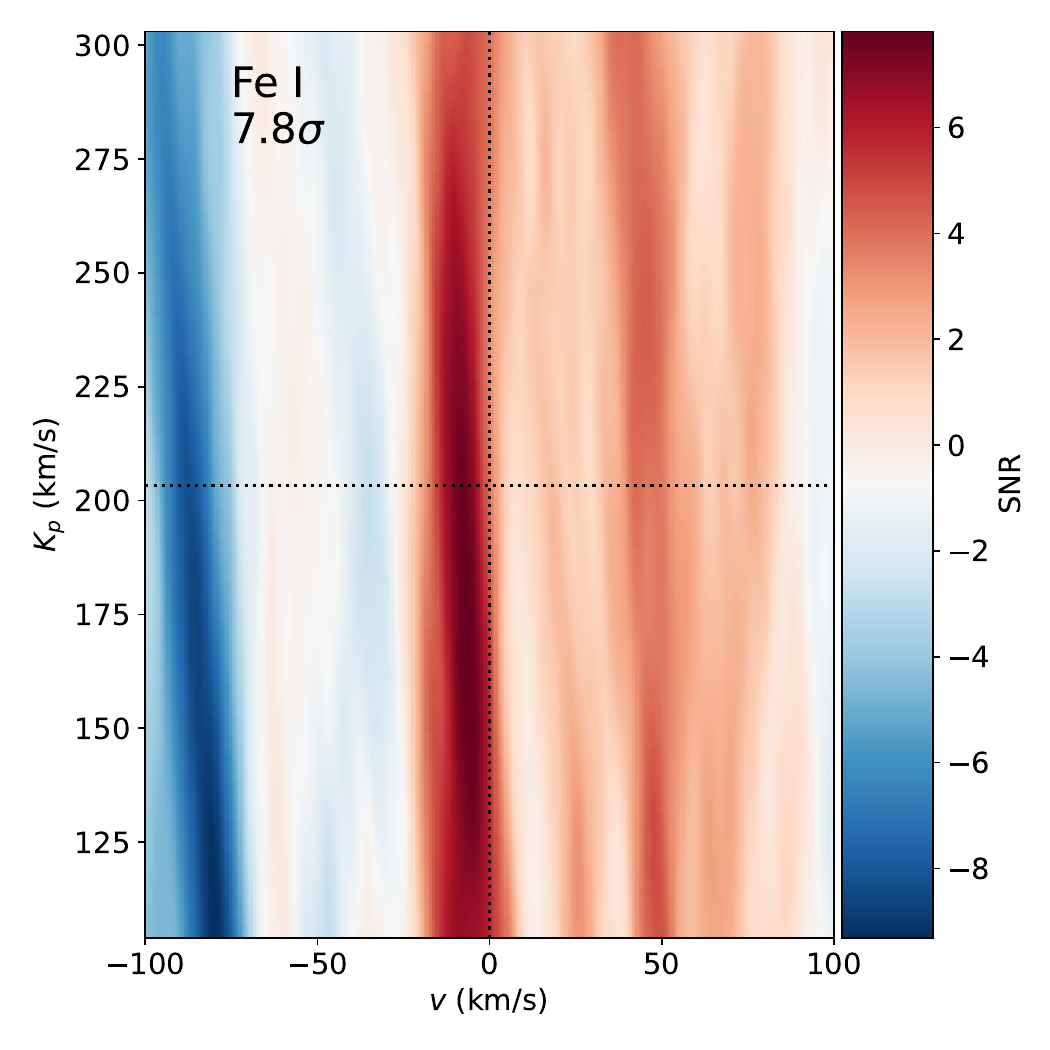}
     \end{subfigure}
     \hfill
     \begin{subfigure}[b]{\columnwidth}
         \centering
         \includegraphics[width=\textwidth]{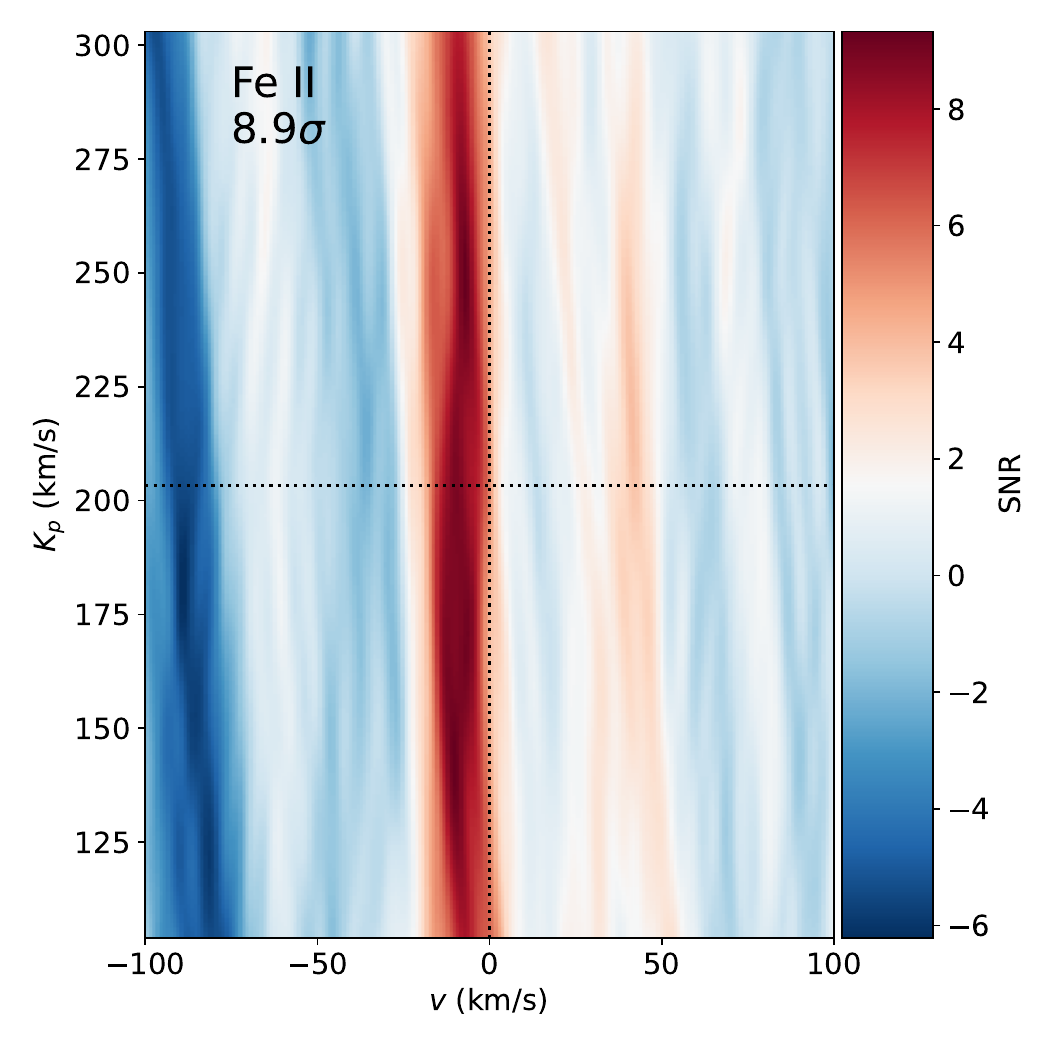}
     \end{subfigure}
     \hfill
     \begin{subfigure}[b]{\columnwidth}
         \centering
         \includegraphics[width=\textwidth]{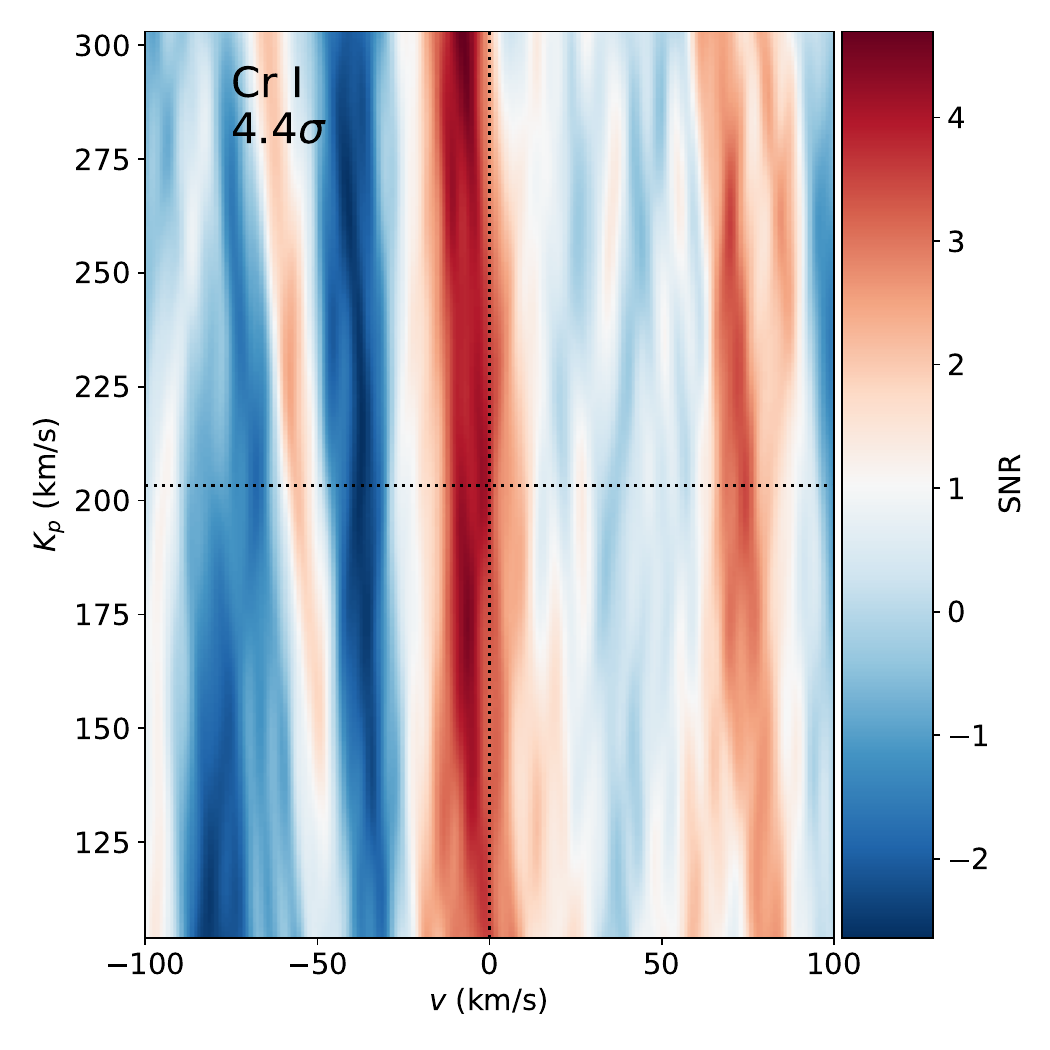}
     \end{subfigure}
     \hfill
     \begin{subfigure}[b]{\columnwidth}
         \centering
         \includegraphics[width=\textwidth]{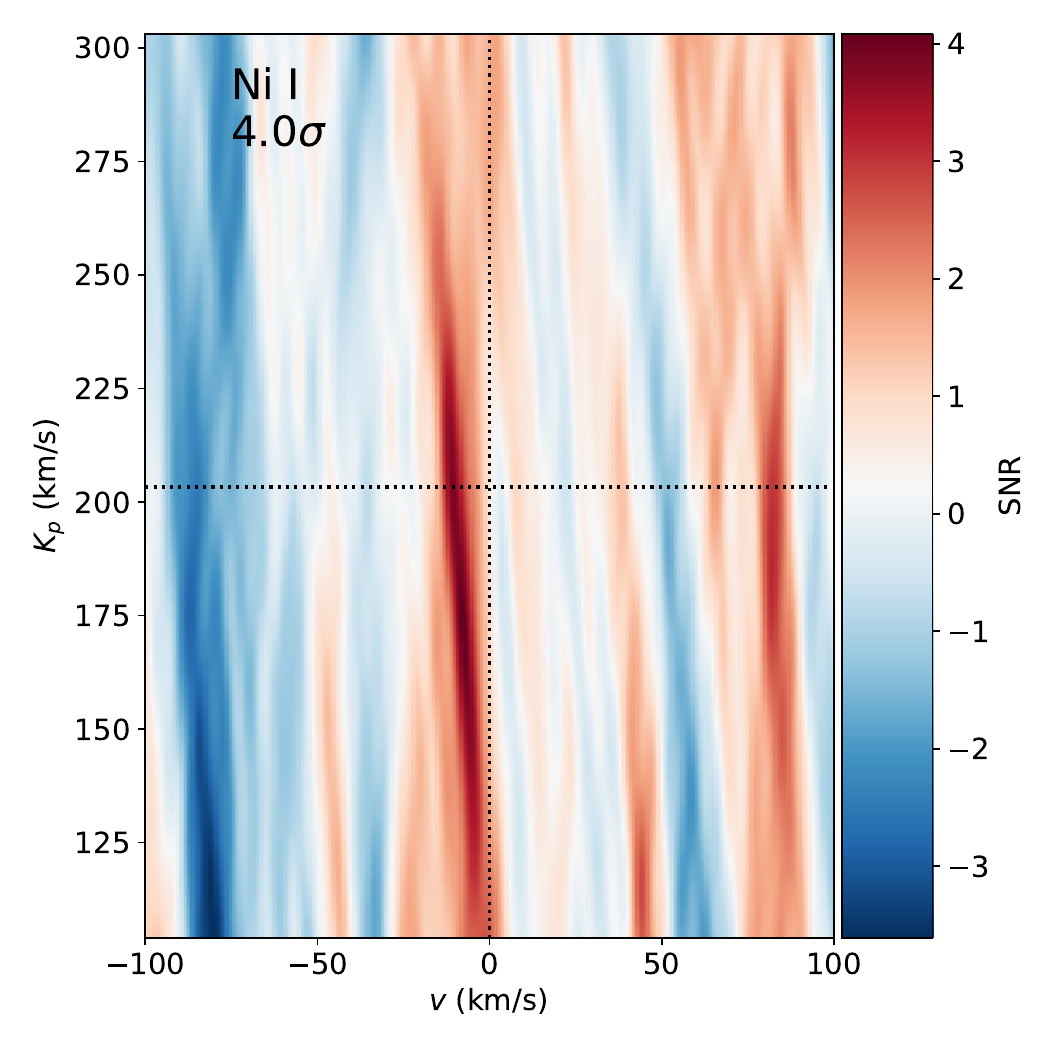}
     \end{subfigure}
        \caption{CCFs shifted into the planetary rest frame over a range of $K_P$ values for our detected species. For each species, the red and blue arm data was combined using the weights described in \S4. The crosshairs intersect at $v = 0$ and the expected $K_P$ value. Each of these species shows a peak (red) near the crosshairs, with the peak SNR identified in the upper left.}
        \label{fig:ShiftedCCFs}
\end{figure*}

We combine the results from both the blue and the red arms using a weighted sum to calculate the shifted CCF in the upper left panel of Fig.~\ref{fig:ShiftedCCFs}. The weight is the SNR of the spectrum squared multiplied by the total equivalent width of lines in the template spectrum. This weighting scheme accounts for both observational noise from our spectra, as well as the strength of the signal expected for the red arm and the blue arm based on the lines in our templates. We iterate over a grid of radial velocity semi-amplitudes ($K_P$) to shift the CCFs into the planetary rest frame. 
The presence of Fe I is suggested by the red region centered on the crosshairs. The peak SNR is measured as the maximum SNR in a box within $\pm 20$km s$^{-1}$ of these crosshairs. 

\section{Results}

\subsection{Detecting Species}
We repeat the process for several neutral atomic species (Ca, Cr, Ti, Ni), ionized species (Fe II), metal hydrides (NaH, MgH, CaH), and metal oxides (TiO, VO). Our detections and tentative detections are shown in Table~\ref{tbl:species} and Fig.~\ref{fig:ShiftedCCFs}. 

\begin{table}
\centering
\caption{Detected species and $T_{23}$ Winds}
\begin{tabular}{l l l l}
\multicolumn{1}{c}{Species}
&\multicolumn{1}{c}{VMR}
&\multicolumn{1}{c}{SNR}
&\multicolumn{1}{c}{Wind}\\
\multicolumn{1}{c}{}
&\multicolumn{1}{c}{}
&\multicolumn{1}{c}{}
&\multicolumn{1}{c}{[km s$^{-1}$]}\\
\hline
Fe I & $5.4 \times 10^{-4}$ & 7.8 & $-7.9 \pm 1.0$ \\
Fe II & $5.4 \times 10^{-3}$ & 8.9 & $-6.8 \pm 1.0$ \\
Cr I & $7.4 \times 10^{-7}$ & 4.4 & $-8.8 \pm 2.1$ \\
Ni I & $2.8 \times 10^{-7}$ & 4.0 & $-8.0 \pm 1.0$ \\
\end{tabular}
\label{tbl:species}
\end{table}

We define tentative detections as detections having peak SNR $>3\sigma$ and good detections as detections with peak SNR $>5\sigma$. Under this definition, we detected Fe I and Fe II using the combined blue and red arm data (Table~\ref{tbl:species}). The corresponding shifted CCF plots are labeled in the upper panels of Fig.~\ref{fig:ShiftedCCFs}. We also tentatively detected Cr I and Ni I, with their shifted CCF plots shown in the lower panels of Fig.~\ref{fig:ShiftedCCFs}.

\cite{Cabot2021} was the first to detect Fe I at 5.2$\sigma$ and also found evidence of Fe II at 3.4$\sigma$. In comparison, we detect Fe I at 7.8$\sigma$ and Fe II at 8.9$\sigma$, and tentatively detect Cr I at 4.4$\sigma$ and Ni I at 4.0$\sigma$.
Cabot used EXPRES 
on the 4.3 meter Lowell Discovery Telescope. The larger collecting area on the 2x8.4 meter LBT accounts for our higher SNR and the additional detected species.
In addition, the different wavelength ranges observed by each instrument should make the different instruments more/less sensitive to different species \citep{Petz2024PETS}. 

\subsection{Global Day-to-Nightside Winds}
Because transmission spectroscopy investigates the terminator, we are able to measure day-to-nightside wind velocities driven by the temperature contrast between the day and night sides of the exoplanet \citep{Ehrenreich2020,PaiAsnodkar2022}. In Table~\ref{tbl:species}, we present day-to-nightside wind velocities for our species that have at least a tentative detection. These velocities are measured only from observations taken during the full transit depth ($T_{23}$) by selecting the horizontal slice of the shifted CCF plots (Fig.~\ref{fig:ShiftedCCFs}) that has the highest SNR near the expected $K_P$ value. The expected $K_P$ value is calculated using the mass of the star and exoplanet from \cite{Cabot2021} and the inclination from MISTTBORN. 
We then use a Gaussian Process (GP) with MCMC analysis to fit a Gaussian to the array of SNRs corresponding to that horizontal slice. We chose this GP + MCMC analysis to account for oversampling of the instrumental line spread function and correlated noise in our data when determining wind velocities and estimating their uncertainties. 
We adopt the Radial Basis Function (RBF) kernel for the GP. There are five free parameters in this analysis: (1) amplitude of the Gaussian, (2) center of the Gaussian, (3) width of the Gaussian, (4) GP amplitude and (5) GP length scale. We randomly draw 50 samples of the posterior distribution, then determine the velocity that corresponds to the SNR peak of each model. The mean and standard deviations of these velocities are adopted as the day-to-nightside wind velocity and its uncertainty, with a minimum uncertainty of 1 km s$^{-1}$. The choice of a 1 km s$^{-1}$ lower limit for the uncertainty is consistent with the wind velocity measurement uncertainty using PEPSI for other UHJs (e.g., \citealt{PaiAsnodkar2022}).
The significant blueshift is also consistent with a visual inspection of Fig.~\ref{fig:ShiftedCCFs}, comparing the offset between the peak SNR and the vertical line showing $v$ of 0 km s$^{-1}$. 

The wind velocities are determined independently for the data in the blue and red arms. 
Fe I and Fe II have detections in both arms. Cr I and Ni I visually only have detections in one of the two arms, so wind velocities are only reported using the data from these arms. 
We find values ranging from -6.8 to -8.8 km s$^{-1}$. In the literature, \cite{PaiAsnodkar2022} found day-to-nightside winds of $\sim$10 km s$^{-1}$ for KELT-9 b (which has an equilibrium temperature, $T_{eq} \sim 4050$ K; \citealt{Gaudi2017}). \cite{PaiAsnodkar2022} and \cite{CasasayasBarris2020} found winds on KELT-20 b of -2.55 and -2.8 km s$^{-1}$, respectively ($T_{eq} \sim 2262$ K; \citealt{Lund2017}). \cite{Seidel2021} detected winds in the lower atmosphere of WASP-76 b of $\sim$5.5 km s$^{-1}$, similar to the 5.3 km s$^{-1}$ reported by \cite{Ehrenreich2020} (Teq 2228 K; \citealt{Ehrenreich2020}). Our measurements for TOI-1518 b are similar in magnitude to KELT-9 b, despite having a lower $T_{eq} \sim 2492$ K (\citealt{Cabot2021}). Our measurements are larger compared to KELT-20 b and WASP-76 b despite only being marginally hotter. 

\subsection{Time-Resolved Wind Velocities}

Wind velocities are not constant as an exoplanet transits in front of its host star. The first major component of the wind velocities is a blueshifted global wind due to the temperature difference between the day and night sides of the exoplanet. The second major component is the velocity of the equatorial jet, blueshifted on one side of the exoplanet and redshifted on the other. As the leading terminator of the exoplanet begins ingress, the velocity of the equatorial jet on the side of the exoplanet transiting the star shows a redshift. Adding this component to the blueshifted global wind at this point in time results in a lower measured wind speed. Once the other side of the exoplanet begins to transit, the two components combine for a higher measured wind speed.

We search for this effect by creating a phase-binned CCF plot. We first bin our observations based on orbital phase, choosing an initial bin size of 0.005. The aim is to have sufficient SNR in each bin to detect the time-resolved wind velocities, while also having a meaningful number of phase bins to observe changes over the course of the transit. The CCFs in each bin are summed using the CCF weights described in \S4. 
We fit a Gaussian to each bin using the combined GP + MCMC analysis described in \S5.2. We further check that the peak SNR > 3 or that the amplitude of these Gaussians show a SNR > 3. If they do not, we combine the data in this bin with the next bin forward in time and continue the analysis. This gives us a variable bin size which is increased during times of the transit where we have less available data or the data has lower SNR. We run tests in binning the data with forward and reverse time and confirm that the binning strategy does not affect the conclusions in this paper.

\begin{figure}
     \centering
     \begin{subfigure}[b]{\columnwidth}
         \centering
         \includegraphics[width=\columnwidth]{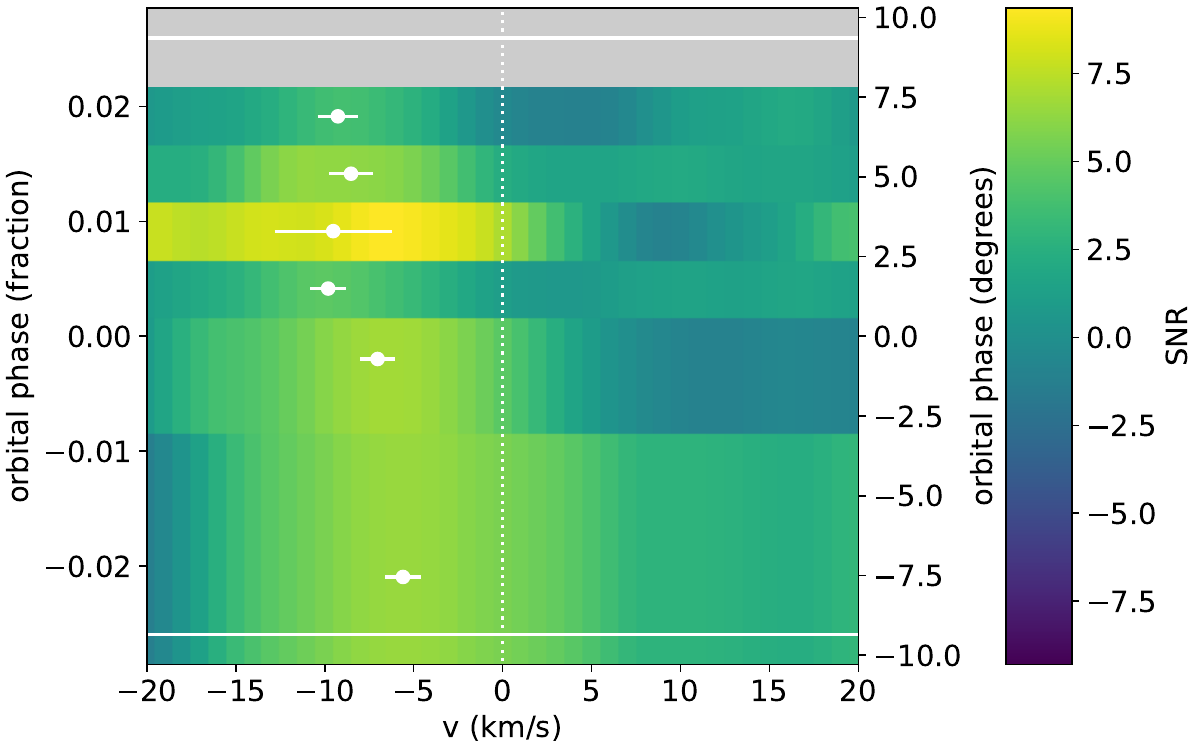}
     \end{subfigure}
     \hfill
     \begin{subfigure}[b]{\columnwidth}
         \centering
         \includegraphics[width=\columnwidth]{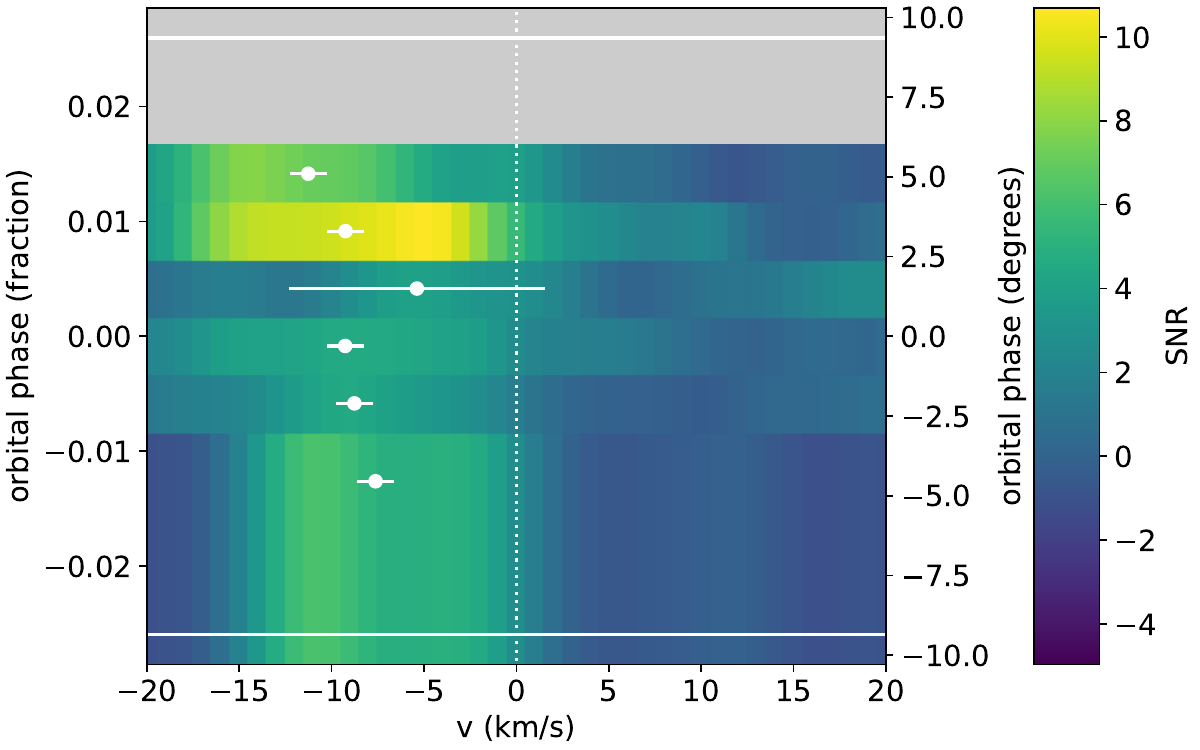}
     \end{subfigure}
     \hfill
        \caption{(Top) Fe I phase bins with wind velocities. (Bottom) Same for Fe II. SNR is shown by color with yellow representing higher SNR. The horizontal white lines show ingress and egress, and the vertical dotted line shows a velocity of 0 km s$^{-1}$. The white points show the wind velocity for each phase bin, fit by a Gaussian using MCMC + GP. The uncertainties are equal to the uncertainties in the centers of the Gaussians, with a minimum uncertainty of 1 km s$^{-1}$. We used dynamic binning, where if the data in one phase bin resulted in a low SNR fit, it was combined with the data in the next phase bin and fitted again. For combined bins, the orbital phase is the weighted sum of each orbital phase bin weighted by their SNR (amplitude of the Gaussian). We neglect original bins with negative Gaussian amplitudes due to weak SNR.}
        \label{fig:PhaseBinnedCCFs}
\end{figure}

Fig.~\ref{fig:PhaseBinnedCCFs} shows the phase binned plots with the wind velocities. 
The velocities of the white dots and their errors are the wind velocities adopted from our MCMC + GP analysis. 
For Fe I (top), the mean wind velocities decrease slightly over the course of the transit. The scatter in the later points is consistent with a constant blueshifted value as was observed in WASP-76~b \citep{Ehrenreich2020}. However, the pattern observed for WASP-76~b also showed a clear increase in velocity after the start of transit. While we see a small increase in velocity, our missing observations near ingress prevent us from drawing a definitive conclusion that we observe the same pattern over the entire course of the transit. We measured a similar pattern of wind velocities for Fe II as for Fe I. 

We compute a reduced $\chi^2$ to determine whether the wind velocities measured for Fe I and Fe II are consistent with each other. This is computed as 
\begin{equation}
    \chi^2 = \sum_i \frac {(RV_{1,i}-RV_{2,i})^2} {\sigma_{RV_{1,i}}^2+\sigma_{RV_{2,i}}^2}
\end{equation}
divided by $\nu = 6$ degrees of freedom (dof). Where there were missing values in one of the phase bins, we used linear interpolation to find the wind velocity and uncertainty from the nearest points. For the 6 data points of Fe I (Fe II), this gives a value of 1.3 (1.6). For 6 dof, a value $>$ 2.1 would reject the null hypothesis at a 95\% confidence level (see, e.g., \citealt{Pugh1966}). This suggests that the wind velocities we measured are consistent with coming from the same underlying distribution.

Considering that we are missing some observations near ingress, our measurements of the time resolved wind velocities could be consistent anywhere from a similar interpretation as the winds in WASP-76~b \citep{Ehrenreich2020} to a constant wind velocity. We perform a statistical test to check if the data prefers (1) a vertical constant line model or (2) a toy model with a linear change in velocity followed by a constant velocity. The second model would be consistent with the physical interpretation from~\citet{Ehrenreich2020}. For Fe I, the reduced $\chi^2$ values are 1.6 and 2.3 for Model \#1 and \#2. For Fe II, the reduced $\chi^2$ values are 1.0 and 0.7 for Model \#1 and \#2. Based on these values and the uncertainty of error bars, we cannot draw a strong conclusion as to whether our data prefers a vertical constant model or a linear+constant toy model. The SNR for Cr I and Ni I is not high enough to detect the time-resolved wind velocities in smaller phase bins.

\subsection{Nodal Precession}
The nodal precession of TOI-1518~b, with a rate of change in the impact parameter $db/dt = -0.0116 \pm 0.0036 \text{ yr}^{-1}$, was discovered in \cite{Watanabe2024}. Combined with their impact parameter $b = 0.91497$ from 2019 TESS data, we expect an impact parameter $b \sim 0.890 \pm 0.0077$ during our 2022 September 26 epoch. Note that the best fit line to the data points in \cite{Watanabe2024} passes slightly under the 2019 impact parameter value, so this expected value is slightly too high. Our measurement of the impact parameter would fit within $1\sigma$ of the best fit line in their work. Other transit parameters from MISTTBORN are generally similar to the values in previous works. 
One notable difference is that we find a smaller ratio of planetary to stellar radius, but this was expected due to a scale factor that was applied to the data in MISTTBORN to match the amplitude of the Doppler shadow to the data. A more thorough analysis of nodal precession will be completed in a future paper.

\section{Conclusions}

We used transmission spectra from the PEPSI spectrograph on the LBT and generated template spectra using petitRADTRANS, then computed CCFs to search for species in the atmosphere of TOI-1518~b. We detected Fe I (7.8$\sigma$) and Fe II (8.9$\sigma$) given our detection threshold of 5$\sigma$, and tentatively detected Cr I (4.4$\sigma$) and Ni I (4.0$\sigma$). This confirms the Fe I and Fe II detections in previous works (\citealt{Cabot2021} in transmission, \citealt{Petz2024PIRANGA} in emission), and \cite{Petz2024PIRANGA} also tentatively detected Ni I in emission.

We further placed our CCFs into bins as a function of orbital phase to detect the wind velocities of Fe I and Fe II over time. Cr I and Ni I did not have sufficient SNR to reliably detect a signal in most phase bins. We used an MCMC + GP analysis to fit a Gaussian to the SNR for each phase bin. The results are shown in Fig.~\ref{fig:PhaseBinnedCCFs}. For Fe I, we observed a slight increase in blueshift until some constant velocity was reached, similar to the results in \cite{Ehrenreich2020} and \cite{Borsa2021}. However, our missing observations near ingress prevented us from definitively resolving the expected velocity change. The wind velocities of Fe II are consistent with coming from the same underlying distribution as Fe I. 

Future studies could look to study emission spectra of TOI-1518~b \citep{Petz2024PIRANGA} in conjunction with transmission spectra to fully probe the atmospheric dynamics of the exoplanet. Additional transmission spectra near ingress could also fill in the gaps in our data caused by weather, allowing higher SNR detections of the winds at the beginning of the transit. 

The authors have been made aware that during the course of writing this paper, another paper by \cite{Simonnin2024} has appeared on arXiv with a complementary analysis of TOI-1518~b. They used MAROON-X to detect 14 species in the atmosphere of TOI-1518~b including Fe I, Fe II, and Cr I as identified in this paper, but not including Ni I. Their detection of Ti I is also of interest as an alternative to detecting TiO, which has been elusive to date despite being a candidate to cause temperature inversions. \cite{Simonnin2024} also present 6 of these species (Fe I, Fe II, Ca I, Ca II, Na I, and Mg I) with time-resolved wind velocities. The velocity patterns of Fe I and Fe II are qualitatively consistent between the two papers, with a constant or slightly increasing blueshift over part of the transit. However, our missing or poor quality observations during ingress preclude a full comparison.

\section*{Acknowledgements}

MCJ, JW, and CL were supported by NASA Grant 80NSSC23K0692. This research has made use of the NASA Exoplanet Archive, which is operated by the California Institute of Technology, under contract with the National Aeronautics and Space Administration under the Exoplanet Exploration Program.

The LBT is an international collaboration among institutions in the United States, Italy and Germany. LBT Corporation Members are: The Ohio State University, representing OSU, University of Notre Dame, University of Minnesota and University of Virginia; LBT Beteiligungsgesellschaft, Germany, representing the Max-Planck Society, The Leibniz Institute for Astrophysics Potsdam, and Heidelberg University; The University of Arizona on behalf of the Arizona Board of Regents; and the Istituto Nazionale di Astrofisica, Italy. Observations have benefited from the use of ALTA Center (\url{alta.arcetri.inaf.it}) forecasts performed with the Astro-Meso-Nh model. Initialization data of the ALTA automatic forecast system come from the General Circulation Model (HRES) of the European Centre for Medium Range Weather Forecasts.

%%%%%%%%%%%%%%%%%%%%%%%%%%%%%%%%%%%%%%%%%%%%%%%%%%
\section*{Data Availability}

Our PEPSI data will be made available through the NASA Exoplanet Archive: \href{https://exoplanetarchive.ipac.caltech.edu/docs/PEPSIMission.html}{https://exoplanetarchive.ipac.caltech.edu/docs/PEPSIMission.html}.



\bibliographystyle{mnras}
\bibliography{references} % if your bibtex file is called example.bib

%\begin{thebibliography}{99}
%\end{thebibliography}

%%%%%%%%%%%%%%%%%%%%%%%%%%%%%%%%%%%%%%%%%%%%%%%%%%

%%%%%%%%%%%%%%%%% APPENDICES %%%%%%%%%%%%%%%%%%%%%

%%%%%%%%%%%%%%%%%%%%%%%%%%%%%%%%%%%%%%%%%%%%%%%%%%

% Don't change these lines
\bsp	% typesetting comment
\label{lastpage}
\end{document}